\documentclass[12pt]{iopart}

\def\>{\rangle}
\def\<{\langle}
\def\ket#1{|#1\>}
\def\bra#1{\<#1|}
\def\braket#1#2{\< #1 | #2 \>}
\def\bracket#1#2#3{\< #1 | #2 | #3\>}
\def\ave#1{\< #1\>}

\newcommand{\op}[1]{\hat{#1}}
\def\rhom{\rho_{{\rm M}\delta}(t)}
\def\rhos{\rho_{{\rm s}\delta}}
\def\F{F(t)}
\def\Fr{F_{\rm R}(t)}
\def\Frr{F_{\rm R}^2(t)}
\def\Fp{F_{\rm P}(t)}
\def\trs{\tr_{\rm s}}
\def\tre{\tr_{\rm e}}
\def\Jy{J_{\rm y}}
\def\Jx{J_{\rm x}}
\def\Jz{J_{\rm z}}
\def\s{\rm s}
\def\e{\rm e}

\usepackage{bbm}
\usepackage{bm}
\usepackage{graphicx}

\begin{document}

\title{Fidelity and Purity Decay in Weakly Coupled Composite Systems}

\author{Marko \v Znidari\v c and Toma\v z Prosen}

\address{Physics Department, Faculty of Mathematics and Physics, 
University of Ljubljana, Jadranska 19, SI-1000 Ljubljana, Slovenia}

\eads{\mailto{znidaricm@fiz.uni-lj.si}, \mailto{prosen@fiz.uni-lj.si}}  

\begin{abstract}
We study the stability of unitary quantum dynamics of composite systems 
(for example: central system + environment) with respect to weak
interaction between the two parts. Unified theoretical formalism is applied to study different physical situations: (i) coherence of a forward evolution as measured by purity of the reduced density matrix, (ii) stability of time evolution with respect to small coupling between subsystems, and (iii) Loschmidt echo measuring dynamical irreversibility. Stability has been measured either by fidelity of pure states of a composite system, or by the so-called {\em reduced fidelity} of reduced density matrices within a subsystem. 
Rigorous inequality among fidelity, reduced-fidelity and purity is proved
and a linear response theory is developed expressing these three quantities 
in terms of time correlation functions of the generator of interaction. 
The qualitatively different cases of regular (integrable) or mixing 
(chaotic in the classical limit) dynamics in each of the subsystems are 
discussed in detail. Theoretical results are demonstrated and confirmed 
in a numerical example of two coupled kicked tops.
\end{abstract}


\section{Introduction}

Recently we have witnessed a strong revival of interest in the theoretical questions 
related to decoherence, in particular due to an immense practical potential of the 
upcoming quantum information processing technology \cite{qcomp}. In order to design quantum 
devices capable of coherent quantum manipulation, one has to be able to control and 
minimize the decoherence due to an unavoidable weak coupling between the system of the 
device and the environment. Traditionally, one uses idealized
harmonic heat bath models of an environment, and very often also harmonic description
of the central system (e.g. \cite{CL}).
However, the rate of decoherence, or quantum dissipation, may depend on the intrinsic
dynamics of the central system, whether be chaotic or not.
It has been argued by Zurek \cite{Zurek91,Zurek&Paz} that the rate of decoherence as characterized
by the Von Neumann entropy of the reduced density matrix increases with the rate which is
given by the classical phase space stretching rate (Lyapunov exponent), for the quantum 
state which is initially given as minimal uncertainty Gaussian wave-packet and for
sufficiently short times.
Recently, Jalabert and Pastawski have studied the so-called {\em quantum Loschmidt
echo} \cite{Jalabert}, or {\em quantum fidelity}, which may be treated as a measure of 
dynamical reversibility under a slight change in the Hamiltonian, and found a similar 
relation to classical stretching rate for short times as for decoherence. 
This reflects correspondence between classical and quantum evolution of 
wave packets up to the Ehrenfest time $-\log{\hbar}$ \cite{Berman}.
Later on, numerous papers appeared addressing related issues \cite{Jacquod,papers}.

On the other hand, we have shown \cite{Prosen01,Ktop} 
that the behaviour of quantum fidelity may be completely
different, if either the time is longer than the Ehrenfest time, or if the initial (pure) state is more complex, e.g. random. In general, the rate of fidelity decay is given by the time-integral of autocorrelation
function of the perturbation operator, and this is bigger the less chaotic is the 
dynamics thus making the fidelity lower, and vice versa.

In previous papers \cite{Prosen01,Ktop,Decoh,Purfid,QC} we have studied fidelity and the so-called
purity fidelity characterizing the stability of quantum evolution, or quantum echoes, 
to the perturbation of dynamics in a generally coupled composite system, 
i.e. where both perturbed and unperturbed systems were coupled. 
In a situation where we are interested only in the properties of a central subsystem, 
one usually does not have the above general situation but a more specific one. 
Namely, the coupling to the environment is usually unwanted and small and so in an ideal 
(unperturbed) case we would like to have two decoupled systems. In that case we are 
interested in how the coupling of a central system to the environment changes the properties
of the central system. So rather than comparing evolution in two general systems, 
in which both have coupling between subsystems, we have a situation where in an 
unperturbed case the systems are uncoupled and become coupled only because of 
the perturbation. This situation is studied in the present paper.

We develop a unified theoretical framework to deal with different physical situations: (i) coherence of forward evolution as measured by purity of the reduced density matrix traced over the subsystem, (ii) stability of time evolution with respect to small coupling between the subsystems, and (iii) Loschmidt echo measuring dynamical irreversibility. Stability is measured either by fidelity of pure states of a composite system, or by the so-called {\em reduced fidelity} of reduced density matrices within a subsystem. We find general linear response formula expressing the fidelity, the reduced fidelity and the purity, in terms of time-correlation functions of the generator of the perturbation within the subsystems. We emphasize that the decay rates as given by  linear response formalism are usually valid also in the regime of small fidelity/purity. Our general qualitative conclusion is that all the three quantities decrease slower with the increasing chaoticity
of the dynamics in the subsystems. Our theoretical results are clearly demonstrated 
on a system of two coupled quantized kicked tops \cite{Miller}. 
In addition we find some intriguing numerical results on algebraic long-time tails of
some of the stability measures.

\section{Characterizing stability and coherence of reduced time evolution}

Our system consists of a central system and an environment, 
henceforth denoted by subscripts ``s'' and ``e'', respectively. The 
Hilbert space of a composite system is a direct product 
${\cal H}={\cal H}_{\s}\otimes {\cal H}_{\e}$. We will 
compare quantum evolutions generated by two Hamiltonians, the unperturbed $H$, and 
the perturbed $H_\delta$,
\begin{equation}
H=H_{\s}\otimes \mathbbm{1}_{\rm e} + \mathbbm{1}_{\rm s}\otimes H_{\e}, \qquad H_\delta=H+\delta \cdot V,
\label{eq:H}
\end{equation}
where $H_{\rm s,e}$ acts only on the corresponding subspace ${\cal H}_{\rm s,e}$ and $\delta$ is a dimensionless coupling strength, and $V$ is a general perturbation operator which couples both systems. 
The usual measure of stability of overall unitary evolution on the total
Hilbert space ${\cal H}$ is the pure state fidelity (equivalent to quantum Loschmidt echo),
which is the overlap between perturbed $\ket{\psi_\delta(t)}=U_\delta(t)\ket{\psi(0)}$ 
and unperturbed $\ket{\psi(t)}=U(t)\ket{\psi(0)}$ time evolving states, 
where $U(t)=\exp(-{\rm i}Ht/\hbar)$ and $U_\delta(t)=\exp(-{\rm i}H_\delta t/\hbar)$ 
are the unitary propagators. It turns useful to define an {\em echo operator} $M_\delta(t)$ \cite{Decoh,Purfid} \footnote{Note that here the order of perturbed and unperturbed propagators is interchanged with respect to refs. \cite{Decoh,Purfid} in order to facilitate exact partial tracing, but is otherwise inessential.} 
\begin{equation}
M_\delta(t):=U^\dagger(t) U_\delta(t).
\label{eq:M}
\end{equation} 
Writing the density matrices, 
\begin{eqnarray}
\rho(t) &:=& \ket{\psi(t)}\bra{\psi(t)} 
= U(t)\rho(0)U^\dagger(t),\\
\rho_\delta(t) &:=& \ket{\psi_\delta(t)}\bra{\psi_\delta(t)} 
= U_\delta(t)\rho(0)U_\delta^\dagger(t),\\
\rhom &:=& M_{\delta}(t) \rho(0) M_\delta^\dagger(t),\label{eq:rhom}
\end{eqnarray}
the fidelity can be concisely written as
\begin{equation}
\F:=|\braket{\psi_\delta(t)}{\psi(t)}|^2={\rm tr}[\rho(0) \rhom].
\label{eq:F_def}
\end{equation}
The echo operator (\ref{eq:M}) can be rewritten as
\begin{equation}
M_\delta(t)=\op{\cal T}\exp\left(-{\rm i}\Sigma(t)\delta/\hbar\right),
\quad{\rm with}\quad
\Sigma(t):=\int_0^t{V(\tau)d\tau},
\label{eq:Mt}
\end{equation}
where $\op{\cal T}$ is a time-ordering operator, 
and $V(t):=U^\dagger(t) V U(t)$ is the perturbation in the interaction picture. 
This representation of the echo operator (\ref{eq:Mt}) is very convenient
as it can be used \cite{Purfid} to derive various results on the
behaviour of fidelity (\ref{eq:F_def}) and related functions.
As $\rhom$ is nothing else but the density operator of the total system in the
interaction picture, it satisfies
\begin{equation}
\frac{d}{dt} \rhom =\frac{\rm i\delta}{\hbar} [\rhom,V(t)],
\label{eq:rhomD}
\end{equation}
where $[A,B]:=AB-BA$. 
\par
The fidelity is a measure of the distance within the whole Hilbert space 
${\cal H}$. On the other hand in the spirit of our study of a central system, 
we would like to have a quantity that would measure distance only within the 
central system space ${\cal H_{\s}}$. So we should not care if the 
environmental states are corrupted by the perturbation, but would just like the 
evolution on the subspace ${\cal H}_{\s}$ to be preserved as closely as possible. 
With this aim we define a new quantity that we call {\em reduced fidelity}, 
which is a fidelity between the reduced perturbed 
$\rhos(t):=\tr_{\e}{\rho_\delta(t)}$ and unperturbed 
$\rho_{\rm s}(t):=\tr_{\e}{\rho(t)}=U_{\s}(t)\rho_{\s}(0) U_{\s}^\dagger(t)$ 
density matrices, which start from the same product (disentangled) initial state 
$\ket{\psi(0)}=\ket{\psi(0)}_{\s} \otimes \ket{\psi(0)}_{\e}$ and we use 
obvious notation $U_{\s}(t):=\exp{(-{\rm i} H_{\s} t/\hbar)}$. 
The reduced fidelity $F_{\rm R}$ therefore reads
\begin{equation}
\Fr:=\trs{[\rho_{\s}(t) \rhos(t) ]}=\trs{[\rho_{\s}(0) \tre{ \{\rhom\}}]}.
\label{eq:Fr_def}
\end{equation}
The reduced fidelity measures the distance between the two reduced density matrices. 
It can be interpreted either as an inner product between two reduced density
operators propagated by two nearby hamiltonians, or as an inner product
between the initial and the final reduced density operator after the echo dynamics.
\par
On the other hand, if we are interested only in the separability (disentanglement) of the 
final echo density matrix $\rhom$, the relevant quantity is {\em purity-fidelity}
\cite{Decoh,Purfid} 
\begin{equation}
\Fp:=\trs{[\tre{\rhom}]^2}.
\label{eq:Fp_def}
\end{equation}
Now we shall use the fact that $H$ is separable in two subsystems 
(\ref{eq:H}), so that purity-fidelity is in this case equal to {\em purity} 
\cite{Zurek91} of the coupled forward time evolution
\begin{equation}
I(t):=\trs{[\tre \rho_\delta(t)]^2} = \trs{[\rhos(t)]^2}.
\label{eq:I_def}
\end{equation}
In order to see that purity-fidelity is equal to purity in this situation we bring separable propagator 
($U=U_{\rm s} \otimes U_{\rm e}$) out of the innermost trace and use cyclic property of a trace opeartion in the definition of 
$\Fp$ (\ref{eq:Fp_def})
\begin{equation}
\Fp=\trs{[\tre{(U^\dagger U_\delta \rho(0) U_\delta^\dagger U)}]^2}=
\trs{[U_{\rm s}^\dagger \tre{\{U_\delta \rho(0) U_\delta^\dagger \}} U_{\rm s}]^2}=I(t).
\end{equation}
This can be understood as a consequence  of the purity being constant during the evolution with 
the separable hamiltonian $H$.
However in the general case, the purity-fidelity is a property
of echo dynamics and is different from the purity of forward dynamics. 
Since in this paper we are interested in the former case 
(\ref{eq:H}), we will from now on use a symbol $I(t)$ instead of $\Fp$. 
\par
Summarizing, all the three quantities, namely $\F,\Fr$ and $I(t)$, measure the stability of a 
composite system to perturbations. The fidelity $\F$ measures the stability of a 
whole state, the reduced fidelity gives the stability on ${\cal H}_{\s}$ subspace 
and the purity measures separability of $\rho_\delta(t)$. One expects that the fidelity 
is the most restrictive quantity of the three - $\rho(t)$ and $\rho_\delta(t)$ must 
be similar for $\F$ to be high. For $\Fr$ to be high, only the reduced density matrices 
$\rho_{\rm s}(t)=\tre{[\rho(t)]}$ and $\rho_{{\rm s}\delta}(t)=\tre{[\rho_\delta(t)]}$ 
must be similar, and finally, purity $I(t)$ 
(\ref{eq:I_def}) is high if only $\rho_\delta(t)$ factorizes. In a previous paper 
\cite{Decoh} we proved an inequality $F^2(t) \le \Fp$. Along the same line one can prove the following general
inequality
\begin{equation}
F^2(t) \le \Frr \le I(t).
\label{eq:ineq}
\end{equation}
{\bf Proof}: Write $\rho(0)=\rho_{\s}\otimes\rho_{\e}$.
Uhlmann's theorem (noncontractivity of fidelity) \cite{Uhlmann76} states for 
any pure states $\rhom$ and $\rho(0)$ that 
\begin{equation}
F(t) = \tr{[\rho(0) \rhom]} \le \trs[ \rho_{\s} \tre\{\rhom \}]=\Fr.
\label{eq:Uhlmann}
\end{equation}
Then, squaring and applying Cauchy-Schwartz inequality 
$|{\rm tr}(A^\dagger B)|^2 \le {\rm tr}(AA^\dagger){\rm tr}(BB^\dagger)$ 
we get $F^2(t) \le F_{\rm R}^2(t) \le \trs[\tre \rhom]^2 = \Fp = I(t)$ with equality being satisfied only in the trivial case of $M_\delta(t) \equiv \mathbbm{1}$, i.e. when $\delta=0$.

\section{Linear response}

Next we proceed by expanding all the three important quantities in powers of the
perturbation strength $\delta$. Although this procedure is very elementary, it 
greatly helps in understanding the behaviour of various measures of stability. 
The lowest order is given by various two-point time correlation functions of the
perturbation and the decay time scale can be 
inferred from the behaviour of time integrals of this correlations. What is more, 
the dependence of this time scale on $\delta$ and $\hbar$ as well as on the dynamics 
of a system (fast correlation decay or absence of correlation decay) is explicit. 
We start by expanding the echo operator
\begin{equation}
M_\delta(t) = \mathbbm{1} - \frac{{\rm i}\delta}{\hbar}\Sigma(t) - \frac{\delta^2}{2\hbar^2}
\op{\cal T}\Sigma^2(t) +\cdots
\label{eq:M2nd}
\end{equation}
The leading order expansion of fidelity (\ref{eq:F_def}) is then
\begin{equation}
\F=1-\left( \frac{\delta}{\hbar} \right)^2 C(t), \qquad C(t):=\ave{\Sigma^2(t)} - \ave{\Sigma(t)}^2
\label{eq:F2nd}
\end{equation}
where $\ave{\bullet}$ denotes an expectation in the product initial state 
$\ket{\psi(0)}=\ket{1,1}$, with the general notation for a complete basis of 
Hilbert space $\ket{i,\nu }:=\ket{i}_{\s} \otimes \ket{\nu}_{\e}$. If explicitly 
written out, the coefficient $C(t)$ is just an {\em integral of autocorrelation 
function} and reads
\begin{equation}
C(t) =\int_0^t{\int_0^t{ \left\{ \ave{V(\xi)V(\zeta)}-\ave{V(\xi)}\ave{V(\zeta)} 
\right\}d\xi d\zeta }}.
\label{eq:Ct}
\end{equation}
Similarly, for the reduced fidelity $\Fr$ we obtain
\begin{eqnarray}
\Fr&=&1-\left( \frac{\delta}{\hbar} \right)^2 \{ C(t)-D(t) \}
\label{eq:Fr2nd} \\
 D(t)&:=& \ave{\Sigma(t)(\rho_{\s}\otimes\mathbbm{1}_{\e})\Sigma(t)} 
- \ave{\Sigma(t)}^2= \sum_{\nu \neq 1} |\bracket{1,\nu}{\Sigma(t)}{1,1}|^2 \nonumber,
\end{eqnarray}
and for the purity $I(t)$
\begin{eqnarray}
I(t)&=&1-2\left( \frac{\delta}{\hbar} \right)^2 \{ C(t)-D(t)-E(t) \}
\label{eq:Fp2nd} \\
 E(t)&:=& \ave{\Sigma(t)(\mathbbm{1}_{\s}\otimes\rho_{\e})\Sigma(t)} - \ave{\Sigma(t)}^2= \sum_{i \neq 1} |\bracket{i,1}{\Sigma(t)}{1,1}|^2 \nonumber.
\end{eqnarray}
\par
So far we have not specified any particular form of the perturbation $V$ yet. 
To facilitate calculations, we now assume the simplest and physically well justified 
{\em product} form of the interaction
\begin{equation}
V:=V_{\s} \otimes V_{\e}.
\label{eq:Vdef}
\end{equation}
This is a very natural choice in the studies of decoherence.
Such is the usual case where one writes 
$V=x_{\s} \otimes F_{\e}$ meaning the coupling of {\em position} times 
{\em force}. Henceforth, the operator $V_{\e}$ will be referred to as ``force'' 
and $V_{\s}$ as ``position''. Following this assumption, the three coefficients 
$C(t)$, $D(t)$ and $E(t)$ can be written out explicitly in terms of separate 
correlation functions over different spaces, namely 
$\ave{\bullet}_{\rm e,s} = \bra{1}_{\rm e,s}\bullet\ket{1}_{\rm e,s}$:
\begin{eqnarray}
\fl C(t)&=&\int_0^t{\int_0^t{ \biggl\{ \ave{V_{\s}(\xi)V_{\s}(\zeta)}_{\s} \ave{V_{\e}(\xi)V_{\e}(\zeta)}_{\e}-\ave{V_{\s}(\xi)}_{\s}\ave{V_{\s}(\zeta)}_{\s} \ave{V_{\e}(\xi)}_{\e}\ave{V_{\e}(\zeta)}_{\e} \biggr\}d\xi d\zeta }} \nonumber \\
\fl D(t)&=& \int_0^t{\int_0^t{ \biggl\{ \ave{V_{\s}(\xi)}_{\s} \ave{V_{\s}(\zeta)}_{\s} \Bigl[  \ave{V_{\e}(\xi)V_{\e}(\zeta)}_{\e}-\ave{V_{\e}(\xi)}_{\e}\ave{V_{\e}(\zeta)}_{\e} \Bigr]  \biggr\}d\xi d\zeta }} \nonumber \\
\fl E(t)&=& \int_0^t{\int_0^t{ \biggl\{ \Bigl[ \ave{V_{\s}(\xi)V_{\s}(\zeta)}_{\s}-\ave{V_{\s}(\xi)}_{\s}\ave{V_{\s}(\zeta)}_{\s} \Bigr] \ave{V_{\e}(\xi)}_{\e} \ave{V_{\e}(\zeta)}_{\e} \biggr\}d\xi d\zeta }}.
\label{eq:CDE}
\end{eqnarray}
The above correlation integrals (\ref{eq:CDE}) are starting point for our theoretical investigations.
In certain situations they can be simplified even further. We will 
study four different regimes in which simplification is possible: (i) mixing regime 
(corresponding to chaotic classical dynamics in both subspaces) in which arbitrary correlation functions appearing in (\ref{eq:CDE}) decay to $0$ and their integrals thus grow as $\propto t$, (ii) regular regime in which due to absence of mixing the whole correlation integral $C(t)$ (or $D(t)$, or $E(t)$) 
grows as $\propto t^2$, and two regimes in which we have a separation of time scales, with the time scale 
of the environment being much shorter than the time scale of the central system. In this case we will work out two different regimes depending on the mixing property of the environment, namely (iii) ``fast mixing'' environment where environmental 
correlations decay; and (iv) ``fast regular'' environment where the environmental 
correlation function has a non-vanishing time average value. The decay of 
fidelity and purity-fidelity in mixing and regular regimes has already been extensively 
discussed \cite{Prosen01,Ktop,Decoh,Purfid}. In this work we do not only discuss a new quantity, namely the reduced fidelity, but also the unperturbed dynamics is now separable and thus not ergodic on the total space.

It is interesting to notice that the general inequality (\ref{eq:ineq}) is clearly
satisfied in our linear response results since the functions $E(t)$ and $D(t)$ are
written in terms of sums of non-negative real numbers [eqs. (\ref{eq:Fr2nd},\ref{eq:Fp2nd})] 
and are hence themselves always non-negative.

We want to stress that all the results of this section can be directly translated to the discrete time case of quantum maps (kicked quantum systems) by simply treating $t$ as integer variable and replacing all the integrals by sums, $\int_0^t \rightarrow \sum_{0}^{t-1}$.

\section{Numerical experiment: two coupled kicked tops}
We will now illuminate and demonstrate our theoretical predictions with a 
numerical example of two coupled kicked tops. In addition, numerical 
simulations will provide us with some insight about the asymptotic behaviour 
for long times and small fidelity/purity. 
Note that here the time is discrete (integer) and measured in the number of
kicking periods (steps). A single kicked top \cite{Haake1} 
has a unitary one step (Floquet) propagator
\begin{equation}
U(\alpha,\gamma)=\exp{(- {\rm i} \gamma \Jy )} \exp{(-{\rm i} \alpha \Jz^2/2J)}.
\label{eq:1ktop}
\end{equation}
The level of chaoticity (e.g. the rate of mixing) of a single top can be varied by varying 
$\alpha$ and the time scale can be influenced by changing the angle $\gamma$. 
In order to be able to study the reduced 
fidelity and the purity we couple two kicked tops, where one top will act as an environment 
an the other as a central system. Two coupled kicked tops \cite{Miller,Ktop} have the 
following unitary one step propagator
\begin{equation}
U_\delta := U_{\s}(\alpha_{\s},\gamma_{\s}) U_{\e}(\alpha_{\e},\gamma_{\e}) 
\exp{(- {\rm i}\delta \cdot V/\hbar )},
\label{eq:Ud}
\end{equation}
with $U_{\rm s,e}$ being propagators for a single kicked top of a system or environment. 
The unperturbed propagator $U:=U_{\delta=0}$ is simply obtained by putting $\delta=0$ 
into above expression (\ref{eq:Ud}).  Perturbation $V=V_{\s} \otimes V_{\e}$ will be 
chosen to be of two different forms: for regimes (i),(ii) and (iii) both $V_{\rm s,e}=\Jz/J$ 
have the same form,
whereas in the regime (iv) we will take $V_{\s}=\Jz/J$ and $V_{\e}=\Jz^2/J^2$. 
The reason for chosing a different form of perturbation in case (iv) is that we
there want the environmential time correlation function to have a non-vanishing 
time-average in order to yield generic results.
In other cases (i-iii) the precise form of the force and position operators
is irrelevant, so $V=\Jz/J$ provides the simplest choice.
The Planck constant is determined by $J$ as $\hbar=1/J$, so that the semiclassical 
limit implies $J \to \infty$. The initial condition will always be a direct product
of coherent states for both tops, $\ket{1,1}=\ket{\varphi_{\s},\vartheta_{\s}} \otimes 
\ket{\varphi_{\e},\vartheta_{\e}}$, with the expansion of SU(2) coherent state 
$\ket{\varphi,\vartheta}$ into eigenbasis $\ket{m}$ of $\Jz$ being
\begin{equation}
\ket{\varphi,\vartheta}=\sum_{m=-J}^{J}{\left( {2J \atop J+m} \right)^{1/2} 
\cos{(\vartheta/2)}^{J+m} \sin{(\vartheta/2)}^{J-m} e^{-i m \varphi} \ket{m}}.
\label{eq:SU2coh}
\end{equation}
Let us now work out the details of the fidelity, reduced fidelity and purity decay 
in all four mentioned regimes. 

\subsection{Mixing regime}
If the combined correlation function decays to zero sufficiently fast, so that its time
integral converges and $C(t) \propto t$ (\ref{eq:Ct}), we can define a kind of
transport coefficient
\begin{equation}
\sigma:=\lim_{t \to \infty}{C(t)/2t}.
\label{eq:sigma}
\end{equation}
Linear response formalism thus gives us initial linear decrease of fidelity. 
On the other hand, the coefficient $D(t)+E(t)$, occurring in the expression for purity, 
is small compared to $C(t)$ \cite{Decoh}, namely 
$\{D(t)+E(t)\}/C(t)\sim 1/N_{\s}+1/N_{\e}$ ($N_{\rm s,e}$ are dimensions of 
${\cal H}_{\rm s,e}$) and the same holds for each individual $E(t)$ or $D(t)$. 
Therefore, up to semiclassically vanishing correction, all the three quantities are
expected to decay on a same time scale. What is more, 
if also multi-time-correlation functions fall off fast \cite{Prosen01,Ktop,Decoh},
then the shape of decay for longer times is exponential 
with the exponent given by the linear response formula,
\begin{equation}
F^2(t) = F_{\rm R}^2(t) = I(t)=\exp{(-2t/\tau_{\rm m})}, \qquad 
\tau_{\rm m}=\frac{\hbar^2}{2\delta^2 \sigma}.
\label{eq:Fem}
\end{equation}
This formula is expected to be valid for times longer than the Ehrenfest time or classical mixing
time, depending on the initial coherent or random state, respectively,
and for sufficiently small $\delta$ such that the value of fidelity/purity is still far above
the saturation for the above mentioned time scale (see \cite{Ktop} for a detailed discussion).
Therefore, if the central system and the environment are both separately mixing 
(by mixing we mean decay of all time correlation functions) the fidelity, reduced fidelity and purity 
decay in the same way. This means that the decay of purity, decay of reduced fidelity and decay of 
fidelity have the same principal mechanism, that is ``corruption'' of the state vector 
as a whole
which gives dominant contribution over more subtle issues of destruction of coherence. Note that stronger mixing (usually connected to stronger chaoticity of a classical system) means smaller correlation integral $\sigma$ and thus slower decay of purity, fidelity and reduced fidelity. This is totally opposite to the very short time behaviour described by Refs.\cite{Zurek91,Zurek&Paz,Miller}. 
\par
\begin{figure}[h]
\centerline{\includegraphics{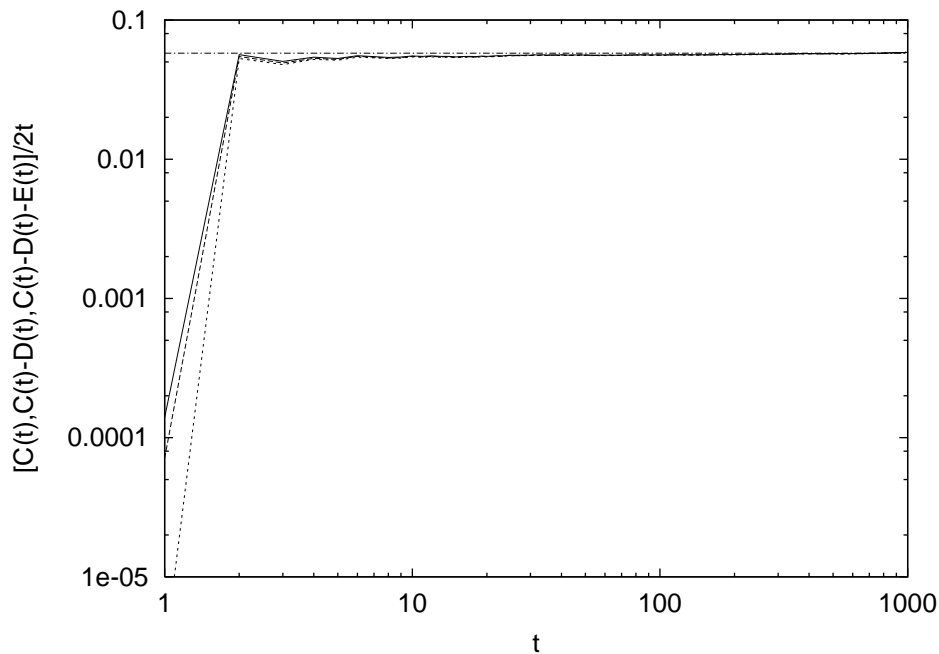}}
\caption{Correlation sums (\ref{eq:CDE}) in the mixing (chaotic) regime (see text) divided by 
$2t$: $C(t)$ (solid), $C(t)-D(t)$ (long dashed),
$C(t)-D(t)-E(t)$ (short dashed). Horizontal (chain) line shows best fitting value of $\sigma$ (\ref{eq:sigma}).}
\label{fig:cor3030}
\end{figure}
\begin{figure}[h]
\centerline{\includegraphics{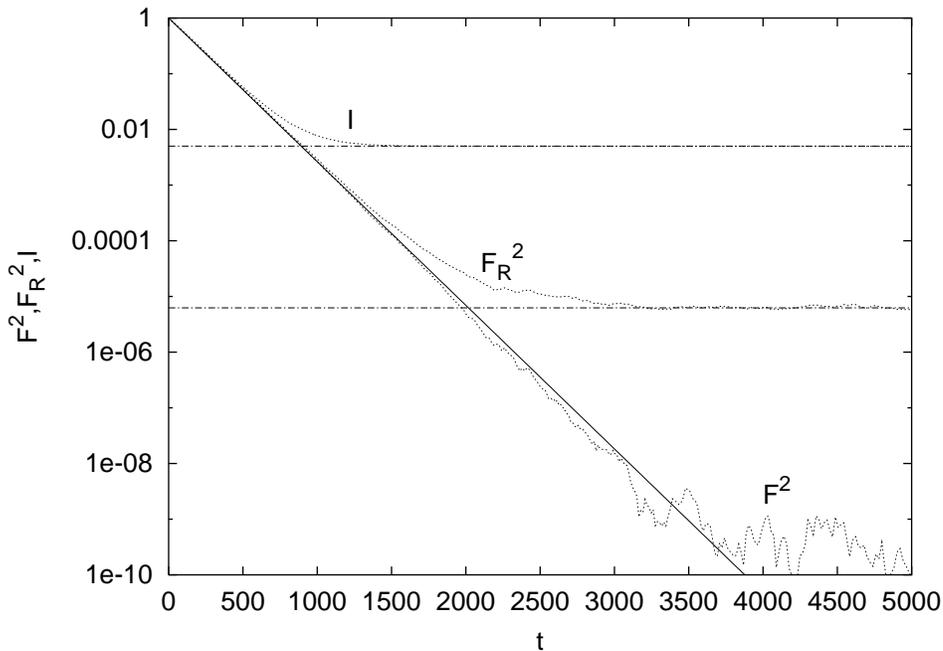}}
\caption{Decay of $F^2(t),\Frr$ and $I(t)$ (dotted curves) in the mixing (chaotic) regime. The solid line gives exponential 
decay (\ref{eq:Fem}) with $\tau_{\rm m}$ calculated from $\sigma$ in figure \ref{fig:cor3030}. 
Horizontal chain lines give the saturation values of purity and reduced fidelity, $1/200$ and $1/400^2$, 
respectively.}
\label{fig:fid3030}
\end{figure}
For numerical verification of this result we chose the perturbation 
$V_{\rm s,e}=\Jz/J$ and parameters $\alpha_{\rm s,e}=30$, $\gamma_{\rm s,e}=\pi/2.1$, and 
$J=1/\hbar=200$ giving the effective size of the Planck constant.
Coherent initial state is placed at $\vartheta_{\rm s,e}=\pi/\sqrt{3}$, 
$\varphi_{\rm s,e}=\pi/\sqrt{2}$. In figure \ref{fig:cor3030} we plot time dependence of 
correlation integrals occurring in expressions for fidelity (\ref{eq:F2nd}), reduced fidelity 
(\ref{eq:Fr2nd}), and purity (\ref{eq:Fp2nd}), showing that the terms $D(t)$ and $E(t)$
are really negligible as all the three quantities shown are practically equal,
for times longer than the Ehrenfest time. 
Next we show in figure \ref{fig:fid3030} the decay of fidelity $F(t)$, reduced fidelity 
$F_{\rm R}(t)$ and purity $I(t)$ for the same parameters and for the perturbation strength 
$\delta=8 \cdot 10^{-4}$. Clean exponential decay is observed in all three cases, 
on a time scale $\tau_{\rm m}$ (\ref{eq:Fem}) given exactly by the lowest order linear response 
expression (\ref{eq:F2nd}) in terms of $\sigma$ (\ref{eq:sigma}) which is obtained from data of 
figure \ref{fig:cor3030}. 
Exponential decay, of course, persists only up to the saturation value determined by a finite 
Hilbert space size \cite{Ktop}.

\subsection{Regular regime}
If the system is regular as a whole, then the time correlation function will generally not 
decay to zero, but will have a non-vanishing average value
\begin{equation}
\overline{c}_{\rm F}:=\lim_{t \to \infty}{C(t)/t^2}.
\label{eq:avgc}
\end{equation}
Similar coefficients can be defined for the average of 
$C(t)-D(t) \asymp \overline{c}_{\rm R} t^2$ occurring in the expansion of the reduced fidelity, 
and $C(t)-D(t)-E(t) \asymp \overline{c}_{\rm I} t^2$ for the purity. Note that $C(t)$ is proportional to 
$\hbar$ for coherent initial states. As shown before \cite{Decoh}, 
the expression $D(t)+E(t)$ is almost equal to $C(t)$ for coherent initial states in the regime
of regular - integrable classical dynamics. 
The expression $C(t)-D(t)-E(t)$ occurring in the formula for purity (\ref{eq:Fp2nd}) is therefore of the
order $\hbar^2$ and the decay time scale for purity decay is $\hbar$ independent. 
This cancellation in the leading order in $\hbar$ happens due to both $E(t)$ and $D(t)$ terms and 
the reduced fidelity will therefore decay on approximately the same time scale as the fidelity. 
For coherent initial states one can 
show \cite{Ktop,Decoh} that the shape of decay is a Gaussian
\begin{equation}
F(t) =\exp{(-(t/\tau_{\rm r})^2)}, \qquad \tau_{\rm r}=\frac{\hbar}{\delta\sqrt{\overline{c}_{\rm F}}},
\label{eq:gauss}
\end{equation} 
and with similar expression, only with $\overline{c}_{\rm R}$ replacing $\overline{c}_{\rm F}$, 
for reduced fidelity. Remember that for coherent initial states $\overline{c}_{\rm F,R} \propto \hbar$ 
and therefore $\tau_{\rm r} \propto \sqrt{\hbar}$. The time scale for decay of purity is again given by 
an analogous expression, namely $\hbar/\delta \sqrt{\overline{c}_{\rm I}}$, with 
$\overline{c}_{\rm I} \propto \hbar^2$, but as we will see in the numerical simulations
and discuss later, its long time behaviour is not a Gaussian but has an algebraic tail 
instead. 
\par
For the numerical demonstration we take two kicked tops with $V_{\rm s,e}=\Jz/J$, $J=200$, 
$\gamma_{\rm s,e}=\pi/2.1$ and $\alpha_{\rm s,e}=0$ in order to have regular dynamics in both subspaces.
Initial coherent state is placed at $(\vartheta,\varphi)_{\s}=(\pi/\sqrt{3},\pi/\sqrt{2})$ and 
$(\vartheta,\varphi)_{\e}=(\pi/\sqrt{3},3\pi/\sqrt{7})$. 
\begin{figure}[h]
\centerline{\includegraphics{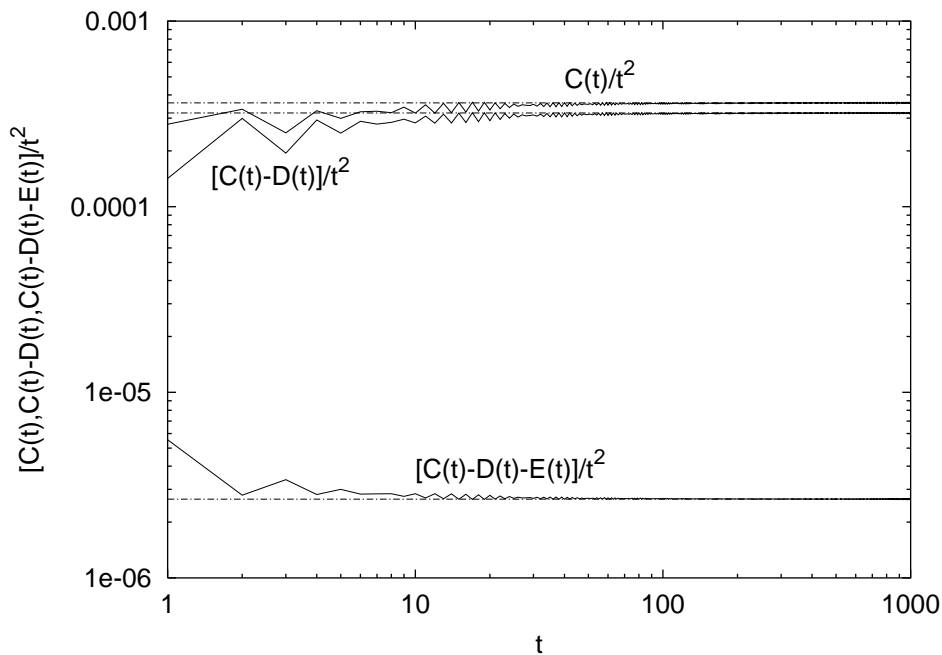}}
\caption{Correlation sums (\ref{eq:CDE}) in the regular regime (solid curves). 
Chain lines show the theoretical values of 
$\overline{c}_{\rm F,R,I}$ (\ref{eq:Thc}). See text for details.}
\label{fig:cor00}
\end{figure}
In fact, the dynamics for $\gamma_{\rm e,s}=\pi/2$ (approximating a 'more generic' value $\gamma=\pi/2.1$, and 
$\alpha=0$) is very simple, namely it is a pure
rotation by an angle $\pi/2$, and all three correlation integrals (sums) can be calculated explicitly. 
Coherent state expectation values of the time averaged correlation 
functions are then easily evaluated using formulas for the expectation values of the lowest powers of 
$\Jx,\Jy,\Jz$. If we denote with $\mathbf{n}_{\rm s,e}=(x,y,z)_{\rm s,e}=
(\sin{\vartheta}\cos{\varphi},\sin{\vartheta}\sin{\varphi},\cos{\vartheta})_{\rm s,e}$ the unit 
vectors in the direction of initial coherent states of the central system and the environment, 
then the results for $\overline{c}$ (for this special case of $\pi/2$ rotation with $\alpha=0$) are
\begin{eqnarray}
\fl \overline{c}_{\rm F}&=&\frac{1}{8J}\left[ 2-y_{\s}^2-y_{\e}^2-2(\mathbf{n}_{\s} \cdot \mathbf{n}_{\e}-y_{\s} y_{\e})^2\right]+\frac{1}{16J^2} \left[ (y_{\s}-y_{\e})^2+(\mathbf{n}_{\s} \cdot \mathbf{n}_{\e}-y_{\s} y_{\e})^2 \right] \nonumber \\
\fl \overline{c}_{\rm R}&=&\frac{1}{8J}\left[ 1-y_{\e}^2-(\mathbf{n}_{\s} \cdot \mathbf{n}_{\e}-y_{\s} y_{\e})^2\right]+\frac{1}{16J^2} \left[ (y_{\s}-y_{\e})^2+(\mathbf{n}_{\s} \cdot \mathbf{n}_{\e}-y_{\s} y_{\e})^2 \right] \nonumber \\
\fl \overline{c}_{\rm I}&=&\frac{1}{16J^2} \left[ (y_{\s}-y_{\e})^2+(\mathbf{n}_{\s} \cdot \mathbf{n}_{\e}-y_{\s} y_{\e})^2 \right]. 
\label{eq:Thc}
\end{eqnarray}
Note that all $\overline{c}$'s are expressed in terms of the invariants of motions as they should be by definition 
(however, this does not mean that the perturbation itself is an invariant of motion). Here one can also explicitly 
see the cancellation of terms for purity, namely the $\overline{c}_{\rm F}$ and $\overline{c}_{\rm R}$ are 
proportional to $\hbar$, while $\overline{c}_{\rm I}$ is proportional to $\hbar^2$. We should note that for 
special positions of initial 
packets, the average correlator may vanish $\overline{c}=0$, and there the decay may be much slower and not Gaussian at all. The zeroes of $\overline{c}$'s are therefore very special points denoting wave packets that are very stable against perturbations. For fidelity and reduced fidelity this slow decay can give rise to a power law decay of an average fidelity (fidelity averaged over the whole phase space). 
\par
In figure \ref{fig:cor00} we can see that the three theoretical coefficients
$\overline{c}$'s (\ref{eq:Thc}) for $\gamma_{\rm s,e}=\pi/2$ approximate very well the 
numerical calculation of correlation sums in the case of $\gamma_{\rm s,e}=\pi/2.1$. 
\begin{figure}[h]
\centerline{\includegraphics{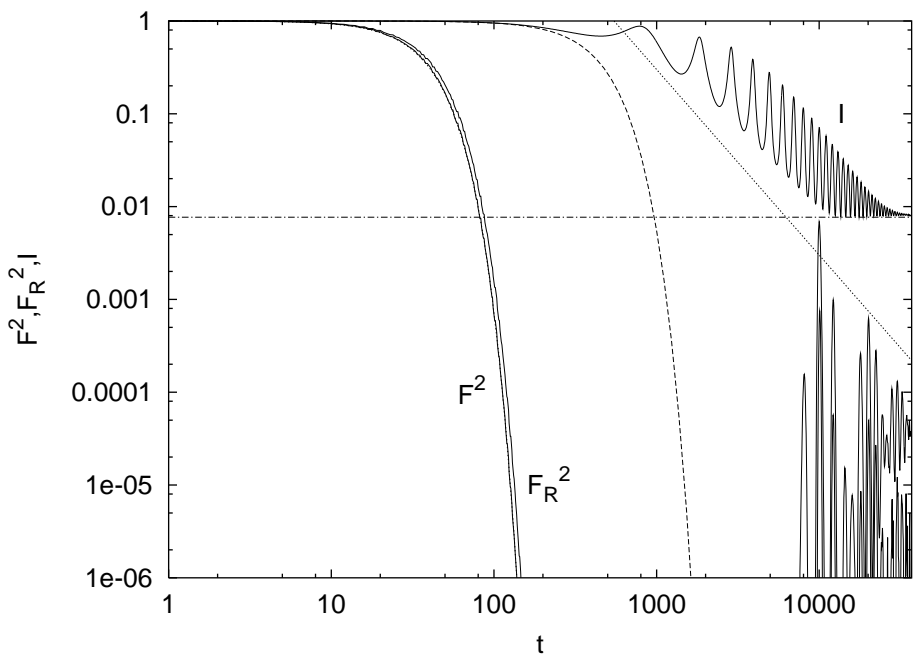}}
\caption{Decay of $F^2(t),\Frr$ and $I(t)$ (solid curves) in a regular regime (see text).
Dashed curve is a Gaussian with the exponent given by $\overline{c}_{\rm I}$. Note that theoretical Gaussians
for $F(t)$ and $F_{\rm R}(t)$ are indistinguishable from the data! Dotted line has a slope -2, and the horizontal 
chain line gives the saturation value of $I(t)$ at $\approx 1/130$.}
\label{fig:fid00}
\end{figure}
In figure \ref{fig:fid00} we show the decay of fidelity, reduced fidelity and purity for $\delta=5 \cdot 10^{-3}$. 
The decay of fidelity and reduced fidelity is Gaussian (\ref{eq:gauss}) with the decay times $\tau_{\rm r}$
given very accurately by the theoretically calculated $\overline{c}_{\rm F}$ and $\overline{c}_{\rm R}$ 
shown with solid 
curves in figure \ref{fig:cor00}. 
The decay of purity on the other hand is not Gaussian for long times. Of course, it decays 
quadratically as given by a linear response formula (\ref{eq:Fp2nd}) for times short enough that the purity is 
close to $1$, but for larger times it decays with a power-law. 
\begin{figure}[h]
\centerline{\includegraphics{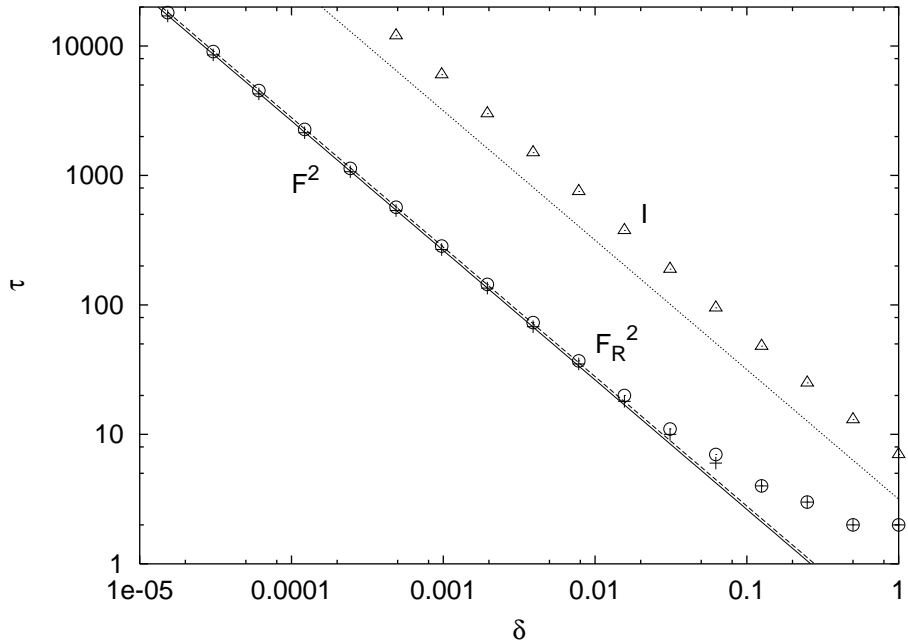}}
\caption{Times $\tau$ at which $F^2(t),\Frr,I(t)$ fall to level $0.37$ for different $\delta$ in a regular regime. 
Symbols: pluses, circles, and triangles, are numerics whereas lines: solid, dashed, dotted, give theoretical dependence of $\tau_{\rm r}$,
for $F^2(t),\Frr$, and $I(t)$, respectively. 
All parameters are the same as for fig. \ref{fig:fid00} and \ref{fig:cor00}, except here: $J=100$.}
\label{fig:presek00}
\end{figure}
Still, up to a constant numerical factor (independent of $\delta$, $\hbar$ etc), 
the effective purity decay time is given by a linear response formula for 
$\tau_{\rm r}$ (\ref{eq:gauss}). This can be observed in figure \ref{fig:presek00} where we show 
numerically calculated values of $\tau_{\rm r}$ based on $1/e \approx 0.37$ level fidelity (purity) 
together with the theoretical prediction on a basis of average correlation function ($\overline{c}$). 
Since the purity decay does not follow a Gaussian
model the numerical and theoretical scale deviate by a numerical factor, which is however constant (independent of
$\hbar$ and $\delta$). For larger perturbation strength $\delta$, the decay rates become less and
less sensitive to perturbations in agreement with the expected saturation \cite{Jalabert,Jacquod}.
\begin{figure}[h]
\centerline{\includegraphics{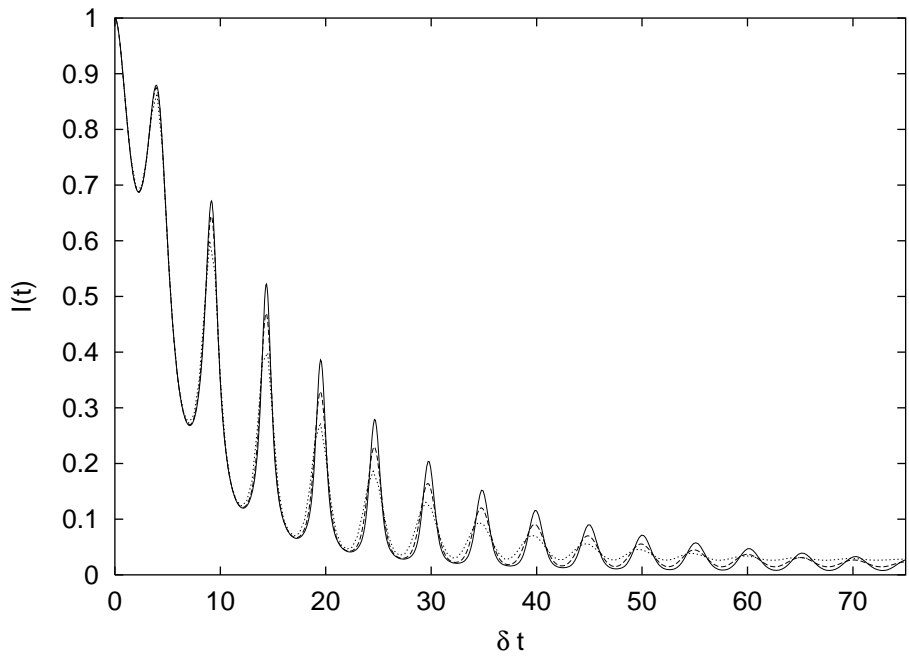}}
\caption{Decay of purity $I(t)$ in a regular regime for $J=200$ (full curve), $J=100$ (dashed) and $J=50$ (dotted) and the same 
parameters as in fig. \ref{fig:fid00}. Note that $I(t)$ is almost fully independent of $\hbar$, and that curves for other values of $\delta$ are completelly overlaping with the existing curves.}
\label{fig:fid0_0_21_21}
\end{figure}
We stress that for short times, namely for $\delta t \ll 1$, the purity of a Gaussian packet for
regular dynamics can be written as an algebraic function in terms of determinants \cite{Decoh}.
This dependence is completely independent of $\hbar$ and is a function of the scaling variable $\delta t$ alone. 
Furthermore, we expect periodic revivals of purity with a classical beating frequency $\nu \sim 1/\delta$, 
again independent of $\hbar$. Based on numerical results (extending the analytical results for longer
times) we conjecture that the overall dependence of $I(t)$ 
is $\hbar$ (or $J$) independent and its functional form can be described in terms of a 
scaling variable $\delta t$, namely $I=f(\delta t)$. This is demonstrated in figure \ref{fig:fid0_0_21_21}. 

The regular regime discussed above has practical importance for the emergence of the macro 
world \cite{Haake&Strunz}. If we have a macroscopic superposition of states the decoherence 
time is going to be smaller than any dynamical time scale in a system and one is trivially in 
a regular regime with non-decaying correlations.

\section{Separation of time scales: Fast environment}
The expressions for $C(t),D(t)$ and $E(t)$ can be further simplified if the decay time scale of the environmental 
correlations $\ave{V_{\e}(t) V_{\e}(t')}_{\e}$ 
is much smaller than the time scale of the systems' correlations 
$\ave{V_{\s}(t) V_{\s}(t')}_{\s}$. 
The time averaging over the fast environmental part of the perturbation $V_{\e}$ can be performed in this case. 
Regarding the environmental correlation function two extreme situations are possible. Namely, the correlations of 
the environment decay (``fast mixing environment'') so that we have a finite integral of environmental 
correlation function, or the correlations of the environment do not decay (``fast regular environment'') and we 
have generically nonvanishing average correlation function of the environment.  

\subsection{Fast mixing environment}
\label{sec:fast_cha}
The situation, when the time scale $t_{\e}$ on which correlation functions for the environment decay
is much smaller than the time scale $t_{\s}$ of the central system, is of considerable physical interest. 
This includes various ``brownian'' like baths, where correlation times are smaller than the dynamical times 
of the system in question. The expressions for $C(t)$, $D(t)$ and $E(t)$ (\ref{eq:CDE}) can be significantly 
simplified in such a situation. We will furthermore assume $\overline{\ave{V_{\e}}}_{\e}=0$, with 
$\overline{\ave{A}}=\lim_{t \to \infty}{t^{-1}\int_0^t{\ave{A(\xi)} d\xi}}$ denoting the time average, which is 
true if we are in an equilibrium situation (average ``force'' $V_{\e}$ vanishes). The integration over the fast 
variable $V_{\e}$ in (\ref{eq:CDE}) can then be carried out and we get 
\begin{eqnarray}
C(t)&=& 2 t \sigma_{\e} \overline{\ave{V_{\s}^2}}_{\s} \nonumber \\
C(t)-D(t)&=& 2 t \sigma_{\e} \left\{ \overline{\ave{V_{\s}^2}}_{\s}-\overline{\ave{V_{\s}}^2}_{\s} \right\}  \nonumber \\
C(t)-D(t)-E(t)&=& 2 t \sigma_{\e} \left\{ \overline{\ave{V_{\s}^2}}_{\s}-\overline{\ave{V_{\s}}^2}_{\s} \right\},
\label{eq:fastCDE}
\end{eqnarray} 
with 
\begin{equation}
\sigma_{\e}:=\lim_{t \to \infty}{\ave{\Sigma_{\e}^2(t)}_{\e}/2t}, \qquad{\rm and}\qquad 
\Sigma_{\e}(t)=\int_0^t{ V_{\e}(\xi) d\xi},
\label{eq:sigmae}
\end{equation}
being the integral of autocorrelation function for the environmental part of perturbation $V_{\e}$ alone. From 
expressions (\ref{eq:fastCDE}) and the lowest order expansions (\ref{eq:F2nd},\ref{eq:Fr2nd},\ref{eq:Fp2nd}) we can 
see that the decay time scale depends only on the time average diagonal correlations of central system 
$\ave{V_{\s}(t)^2}$ and not on the full correlation function. This is a simple consequence of the separation of 
time scales and means that the decay of all the three stability measures 
does not depend on the dynamics of the central system (e.g. being mixing (chaotic) or regular.) 
Furthermore, the reduced fidelity $\Fr$ and the purity $I(t)$ will decay on the same time scale (\ref{eq:fastCDE}), 
meaning that the decay of reduced fidelity is predominantly caused by the loss of coherence, i.e. entanglement between
the two factor spaces.
This means that the reduced fidelity, which is a property of echo dynamics of comparison of two slightly 
different hamiltonian evolutions, is equivalent to the decay of purity or growth of linear entropy of an
individual weakly coupled system.

If the initial state of a central system $\rho_{\s}(0)$ is a Gaussian wave packet (coherent state) then the 
dispersion of $\overline{\ave{V_{\s}^2}}_{\s}-\overline{\ave{V_{\s}}^2}_{\s}$ is by a factor of order 
$1/\hbar$ {\em smaller} than $\overline{\ave{V_{\s}^2}}_{\s}$. Thus for coherent initial states of a central system, no matter what the 
initial state of the environment, the $\Fr$ and $I(t)$ are going to decay on a $1/\hbar$ times {\em longer} time 
scale than $\F$. We have therefore reached a general conclusion based on very weak assumptions of chaotic fast
environment, namely that the 
coherent states are most robust against decoherence (provided $t_{\e} \ll t_{\s}$), and that 
decoherence takes place in times longer than the correlation time of environment $t_{\e} \ll t_{\rm dec}$. If 
decoherence is even faster than the time scale of the environment, as is the case for macroscopic superpositions, 
then formulas (\ref{eq:fastCDE}) are not valid any more as one is effectivelly in a regular regime of the previous section. Decoherence time is then independent not just of systems 
dynamics but also of environmental dynamics characterized by $\sigma_{\e}$ (see \cite{Haake&Strunz}).
\par
\begin{figure}[h]
\centerline{\includegraphics{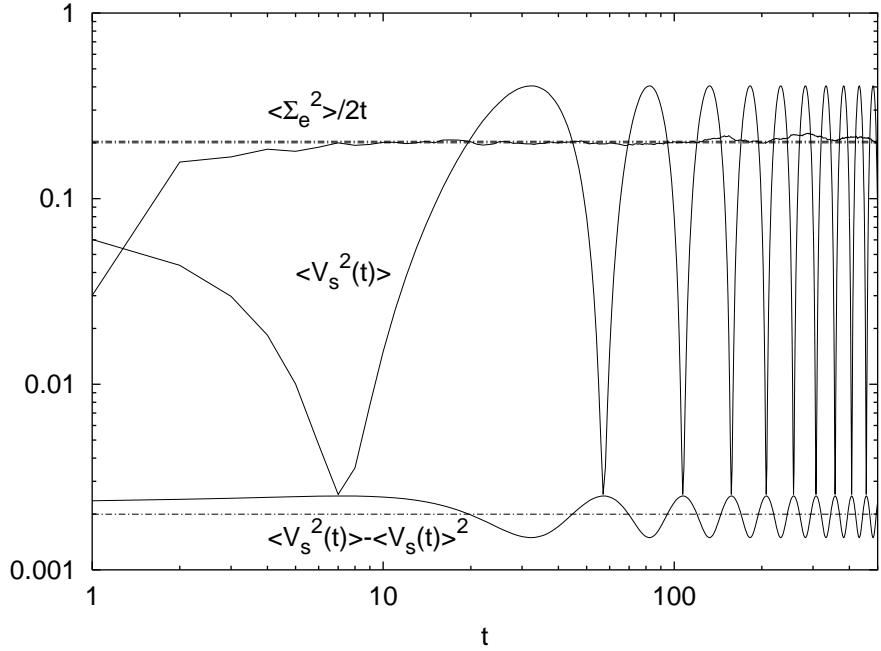}}
\caption{Various correlation sums from formulas (\ref{eq:fastCDE}) in a fast chaotic regime (solid curves, as indicated in the figure). 
Chain lines indicate corresponding theoretical time averages. For details see text.}
\label{fig:cor030}
\end{figure}
In the regime of fast chaotic environment one can immediately derive a master equation for a reduced density 
matrix of a central system \cite{Kolovsky,Qoptics}. We take partial trace over the environment of expansion 
for $\rhom$ (\ref{eq:rhom},\ref{eq:M2nd}) and write it for a small time step $\Delta t$. This time step 
$\Delta t$ must be larger than the correlation time $t_{\e}$ of the environment and at the same time smaller than 
the correlation time $t_{\s}$ of the system. For the environmental part of the correlation function we assume 
fast exponential decay (particular exponential form is not essential) which is independent of the state $\rho$
\begin{equation}
\tr_{\e}{\{V_{\e}(t) V_{\e}(t') \rho \}} \longrightarrow \frac{\sigma_{\e}}{t_{\e}} 
\exp{\{-|t-t'|/t_{\e}\}} \tr_{\e}{\rho}.
\label{eq:expcorr}
\end{equation}
Assuming the perturbation to be a product $V(t)=V_{\s}(t) \otimes V_{\e}(t)$ and the average ``force'' 
$\tr_{\e}{(V_{\e}(t) \rho )}$ to vanish together with the exponential decay of environmental correlations of the 
form (\ref{eq:expcorr}) for arbitrary state, yields a master equation for the reduced density matrix 
$\rho_{\rm Ms}(t):=\tr_{\e}{\rhom}$ 
\begin{equation}
\dot{\rho}_{\rm Ms}(t)=-\frac{\delta^2 \sigma_{\e}}{\hbar^2} [V_{\s}(t),[V_{\s}(t),\rho_{\rm Ms}(t)]].
\label{eq:master}
\end{equation} 
\par
\begin{figure}[h]
\centerline{\includegraphics{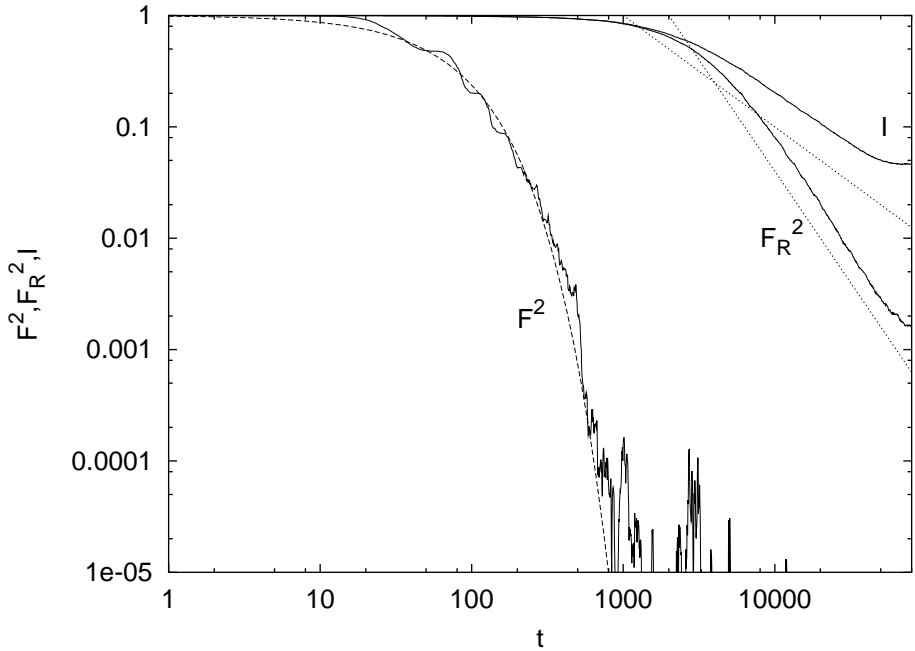}}
\caption{Decay of $F^2(t),F_{\rm R}^2(t)$ 
and $I(t)$ for fast chaotic environment. Dashed line is exponential with the 
exponent given by the values of $\sigma_{\e}$ and $\overline{\ave{V_{\s}^2}}_{\s}$ (\ref{eq:fastCDE}) and two 
dotted lines have slopes $-2$ and $-1$.}
\label{fig:fid030}
\end{figure}
\begin{figure}[h]
\centerline{\includegraphics{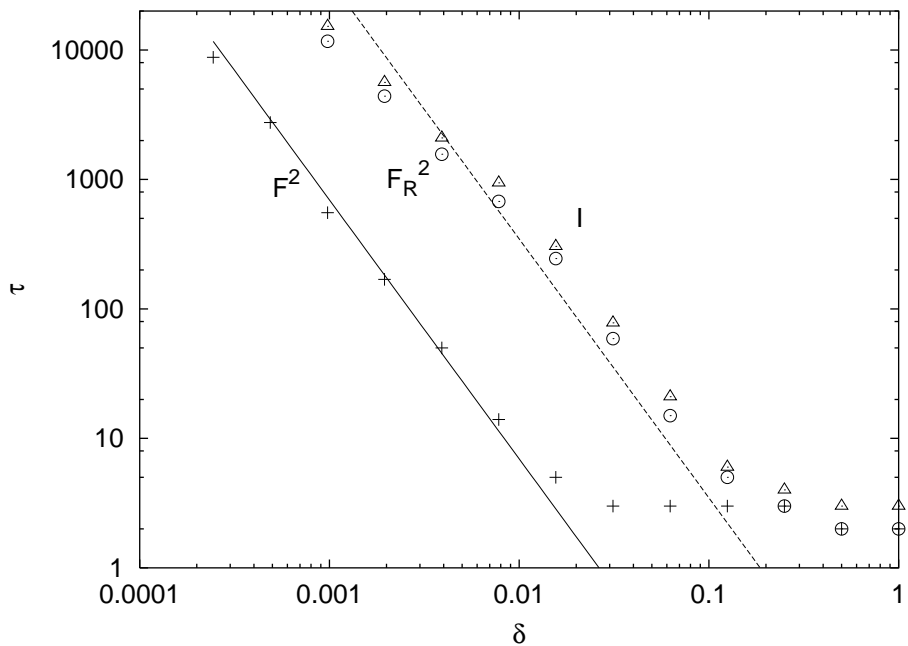}}
\caption{Times $\tau$ at which $F^2(t),F_{\rm R}^2(t),I(t)$
fall to level $0.37$ for different $\delta$ and fast chaotic 
environment. Symbols give numerics and lines give theoretical dependence of $\tau$ (same as in fig.~\ref{fig:presek00}). 
All is for $J=100$.}
\label{fig:presek030}
\end{figure}
For numerical demonstration we choose: $V_{\rm s,e}=\Jz/J$, $J=200$, $\delta=1.5 \cdot 10^{-3}$, coherent initial 
state at $(\vartheta,\varphi)_{\rm s,e}=(\pi/\sqrt{3},\pi/\sqrt{2})$ and parameters $\alpha_{\s}=0$, 
$\gamma_{\s}=\pi/50$ for the central system and $\alpha_{\e}=30$, $\gamma_{\e}=\pi/2.1$ for the environment. 
Actually, we could take any value of $\alpha_{\s}$ and would get qualitatively similar results. The only advantage 
of using regular central dynamics $\alpha_{\s}=0$ is that it is then possible to explicitly calculate averages 
$\overline{\ave{V_{\s}^2}}_{\s}$ 
and $\overline{\ave{V_{\s}^2}}_{\s}-\overline{\ave{V_{\s}}^2}_{\s}$. Namely, if $\alpha_{\s}=0$ and 
$\gamma_{\s} \ll 1$ we get
\begin{eqnarray}
\overline{\ave{V_{\s}^2(t)}}_{\s} &=&\frac{1}{2}(1-y_{\s}^2)+\frac{1}{4J} (1+y_{\s}^2) \nonumber \\
\overline{\ave{V_{\s}^2}}_{\s}-\overline{\ave{V_{\s}}^2}_{\s} &=& \frac{1}{4J} (1+y_{\s}^2).
\label{eq:Vth}
\end{eqnarray}
The values of these two quantities for our initial condition are shown in figure \ref{fig:cor030} with two dotted lines (by pure coincidence we have $\sigma_{\rm e} \approx \overline{\ave{V_{\s}^2(t)}}_{\s}$), together with numerically calculated time dependent (not yet averaged) $\ave{V_{\s}^2(t)}_{\s}$ 
and $\ave{V_{\s}^2(t)}_{\s}-\ave{V_{\s}(t)}^2_{\s}$ for our choice of $\gamma_{\s}=\pi/50$. This time dependent 
values oscillate on a time scale $\approx 50$, which is much longer than the time $\approx 10$ in which 
$\sigma_{\e}$ (\ref{eq:sigmae}) converges and so the assumption $t_{\e} \ll t_{\s}$ is justifiable. The value 
of all these three quantities is then used in linear response formulas (\ref{eq:fastCDE}) to give us time scales 
on which $F,F_R$ and $I$ decay. The results are shown in figure \ref{fig:fid030}. We can see that the fidelity again 
decays exponentially, but the reduced fidelity and purity have a power-law like tails. They decay on a time scale 
still roughly given by the lowest order expansions (\ref{eq:fastCDE}) and the values of $\sigma_{\e}$ (numerical 
from fig. \ref{fig:cor030}) and $\overline{\ave{V_{\s}^2}}_{\s}-\overline{\ave{V_{\s}}^2}_{\s}$ (theoretical 
expression (\ref{eq:Vth})) which can be seen in figure \ref{fig:presek030}. The same general conclussion again holds: the more chaotic the environment is (smaller $\sigma_{\rm e})$, the slower the decay of all three quantities. Purity and reduced fidelity both decay on a $1/\hbar$ longer time scale than the fidelity in accordance with expressions (\ref{eq:Vth}) for coherent initial states. 

\subsection{Fast regular environment}
\label{sec:fast_reg}
Here we will explore perhaps a less physical situation of a regular environmental dynamics.
For a regular environment the double integral of environment correlations grows as $\propto t^2$ due to 
non-decay (plateau) of correlation function and we can define the average correlation function
\begin{equation}
\overline{c}_{\e}:=\lim_{t \to \infty}{\ave{\Sigma_{\e}^2(t)}_{\e}/t^2}.
\label{eq:avgce}
\end{equation}
If in addition the correlations of the system also do not decay then the correlation sum of the total system 
will grow as $\propto t^2$ which is a regular regime discussed previously. 
Here, we will focus on a different situation where the integral of system's correlation function converges, 
that is $C_{\s}(t) \propto t $, i.e. the central dynamics is mixing (chaotic).
We will additionally assume the average ``position'' $V_{\s}$ to be zero $\overline{\ave{V_{\s}}}_{\s}=0$. The 
transport coefficient of a system $\sigma_{\s}$ is then
\begin{equation}
\sigma_{\s}:=\lim_{t \to \infty}{\ave{\Sigma_{\s}^2(t)}_{\s}/2t}, \qquad \Sigma_{\s}(t)=\int_0^t{ V_{\s}(\xi) d\xi}.
\label{eq:sigmas}
\end{equation}
The expressions for $C(t),D(t)$ and $E(t)$ (\ref{eq:CDE}) for the present case can be simplified to
\begin{eqnarray}
C(t)&=& 2 t \sigma_{\s} \overline{c}_{\e}  \nonumber \\
C(t)-D(t)&=& 2 t \sigma_{\s} \overline{c}_{\e}   \nonumber \\
C(t)-D(t)-E(t)&=& 2 t \sigma_{\s} \left\{ \overline{c}_{\e}-\overline{\ave{V_{\e}}}_{\e}^2  \right\}.
\label{eq:fastRegCDE}
\end{eqnarray}
An important thing we notice immediately is that now the reduced fidelity $\Fr$ decays on the same time scale as 
fidelity $F(t)$. This must be contrasted to the case of a fast mixing environment (\ref{eq:fastCDE}), where 
$\Fr$ decayed on the same time scale as purity. If the initial state of the environment $\rho_{\e}(0)$ is a 
coherent state, then the purity will decay on a $1/\hbar$ times longer time scale as fidelity and reduced fidelity.
On the other hand, for a random initial state of the environment, the average force vanishes
$\ave{V_{\e}}_{\e}=0$, and then all the three quantities decay on the same time scale.
\par
\begin{figure}[h]
\centerline{\includegraphics{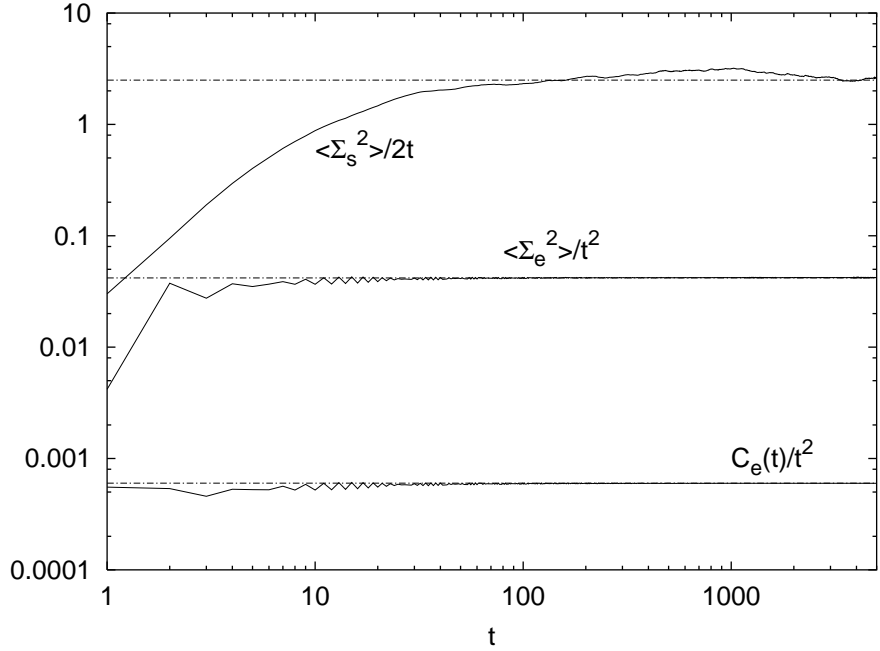}}
\caption{Correlation sums occurring in (\ref{eq:fastRegCDE}) (solid curves). Top chain line gives best fit for $\sigma_{\s}$ and two lower chain 
lines give theoretical time averaged correlation functions for the environment (\ref{eq:Cth}). All is for a fast regular regime. See text for details.}
\label{fig:cor300}
\end{figure}
\begin{figure}[h]
\centerline{\includegraphics{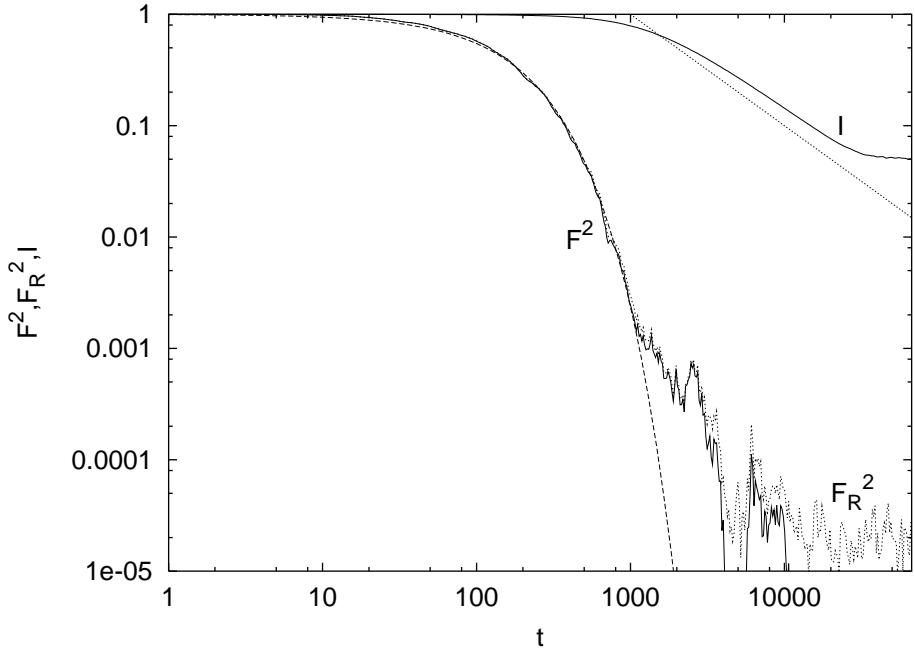}}
\caption{Decay of $F^2(t),\Frr$ and $I(t)$ for fast regular environment. Dashed line is an exponential with the exponent given by a product 
of $\sigma_{\s}$ and $\overline{c}_{\e}$ (\ref{eq:fastRegCDE}). Straight dotted line has a slope $-1$. See text for details.}
\label{fig:fid300}
\end{figure}
\begin{figure}[h]
\centerline{\includegraphics{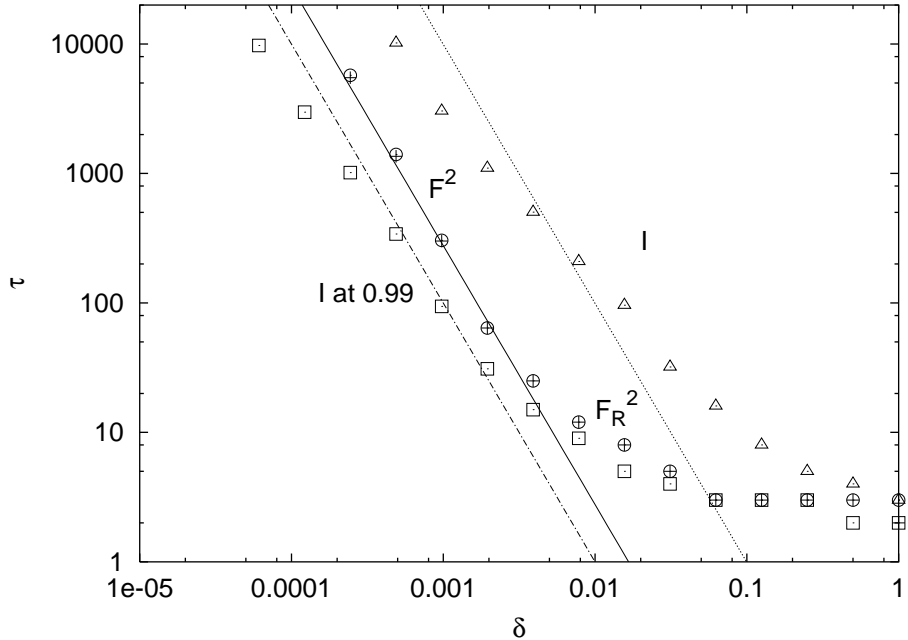}}
\caption{Times $\tau$ at which $F^2(t)$ (pluses), $\Frr$ (circles), $I(t)$ (triangles) fall to level $0.37$, 
and the times $\tau$ when $I(t)$ falls to $0.99$ (squares) for varying $\delta$ and with fast regular environment. Symbols are numerics and lines 
give theoretical dependences of $\tau$. Everything is for $J=100$.}
\label{fig:presek300}
\end{figure}
For the purpose of numerical experiment we chose now $V_{\s}=\Jz/J$, and $V_{\e}=\Jz^2/J^2$ in order to have
a less trivial situation of non-vanishing average force. 
The initial condition is again $(\vartheta,\varphi)_{\rm s,e}=(\pi/\sqrt{3},\pi/\sqrt{2})$ and parameters are 
$J=200$, $\alpha_{\s}=30$, $\gamma_{\s}=\pi/7$ and $\alpha_{\e}=0$, $\gamma_{\e}=\pi/2.1$ and perturbation 
strength $\delta=6 \cdot 10^{-4}$. By choosing explicitly solvable case $\alpha_{\e}=0$ we can calculate $\overline{c}_{\e}$ and $\overline{c}_{\e}-\overline{\ave{V_{\e}}}_{\e}^2$, say for 
the simple case of $\pi/2$ rotation, $\gamma_{\e}=\pi/2$, where we obtain
\begin{eqnarray}
\overline{c}_{\e} &=& \frac{1}{4} (1-y_{\e}^2)^2+\frac{1}{4J}(-3 y_{\e}^4+2 y_{\e}^2 +1) + {\cal O}(1/J^2) \nonumber \\
\overline{c}_{\e}-\overline{\ave{V_{\e}}}_{\e}^2 &=& \frac{1}{2J} y_{\e}^2(1-y_{\e}^2)+\frac{1}{16J^2}(11 y_{\e}^4-11 y_{\e}^2 +2) + {\cal O}(1/J^3).
\label{eq:Cth}
\end{eqnarray}
The values of this coefficients are shown in figure \ref{fig:cor300} (lower two dotted lines). 
They agree nicely with numerics for $\gamma_{\e}=\pi/2.1$. 
In the figure \ref{fig:fid300} we can observe exponential decay of fidelity and reduced fidelity on the same time 
scale (two curves almost overlap) and decay of purity on a $1/\hbar$ longer time scale. For longer times the 
purity decay is again algebraic. In figure \ref{fig:presek300} we show dependence of decay times on $\delta$. 
The dependence for purity is quite interesting. 
If one looks at the time the purity falls to $0.99$ one has the agreement 
with linear response (by definition). But if one looks at the purity level $0.37$, they don't agree as well, meaning
that the nature (shape) of purity decay may change (not only the scale) as one vary $\delta$ or $\hbar$.
On the other hand, this may also be simply a finite size effect due to finite Hilbert space dimensions.

\section{Discussion and conclusions}

In this paper we have analyzed the stability of unitary time evolution of composite systems under the
weak coupling between subsystems. This is a natural (unitary) model for dissipation and decoherence in
quantum mechanics where one subsystem plays the role of the central system and the other plays the role
of the environment. But this is not the only possible application of the above ideas. One may also be
interested in the dynamical effects of weak coupling between two controllable parts of the system, like
e.g. the atom and the electromagnetic cavity \cite{haroche}.

We have analyzed three different quantities which are treated on a similar theoretical footing but which have
different physical interpretations. The first two, namely the fidelity and the reduced fidelity, refer to the
case of echo dynamics where the forward evolution is generated by the uncoupled system while the backward 
evolution is generated by the weakly coupled system. The third one, namely the purity, refers to the
growth of linear entropy or growth of entanglement between the two subsystems during the course of
weakly coupled forward evolution. 
First we have shown a rigorous inequality between the three quantities which may be useful to provide
various bounds.
Then we have developed a linear response theory which predicts time scales for the three quantities
in terms of time correlation functions of the perturbation in each of the subsystems.
Thus we have been able to classify all different behaviours with respect to regularity or chaoticity
(as defined here by the mixing property) of each of the subsystems.
The general conclusion is again, consistent with \cite{Gorin,Tanaka}, that strong chaos stabilizes 
quantum dynamics with respect to the intersystem coupling, and that strong chaos decreases the
rate of entanglement (or linear entropy) growth.
In particular, if the characteristic time scales of correlation decay in two subsystems are well
separated, then we can integrate over the correlation functions of the fast subsystem and obtain 
expressions which are independent of the nature of dynamics in the slow subsystem.

\section*{Acknowledgements}

Useful discussions with T H Seligman are gratefully acknowledged. The work 
has been financially supported by the Ministry of Education, Science and 
Sport of Slovenia, and by the U.S. army research laboratory and the
U.S. army research office under contract n. DAAD 19-02-1-0086.


\section*{References}

\begin{thebibliography}{1}

\bibitem{qcomp} Nielsen M~A and Chuang I~L 2000 {\em Quantum computation and quantum information}~(Cambridge University Press)

\bibitem{CL} Caldeira A O and Leggett A J 1983 {\em Path Integral Approach to Quantum Brownian
Motion} Physica {\bf 121A} 587

\bibitem{Zurek91} Zurek W~H 1991 {\it Decoherence and the transition from quantum to classical} Physics Today {\bf 44} 36

\bibitem{Zurek&Paz} Zurek W H and Paz J P 1995 {\it Quantum chaos: a decoherent definition} Physica {\bf 83D} 300--8

\bibitem{Jalabert} Jalabert R A and Pastawski H M 2001 {\it Environment-independent decoherence rate in classically chaotic
systems} Phys.~Rev.~Lett.~{\bf 86} 2490

\bibitem{Berman} Berman G~P and Zaslavsky G~M 1978 {\it Condition of stochasticity in quantum non-linear systems} Physica {\bf 91A} 450 

\bibitem{Jacquod} Jacquod Ph, Silvestrov P G, Beenakker C W J 2001 
{\it Golden rule decay versus Lyapunov decay of the quantum Loschmidt echo} Phys. Rev. E {\bf 64} 055203(R)

\bibitem{papers} 
Tomsovic S and Cerruti N R 2001 {\it Sensitivity of Wave Field Evolution and Manifold Stability in Chaotic Systems} 
Phys. Rev. Lett. {\bf 88} 054103\\
Cucchietti F M, Pastawski H M and Wisniacki D A 2002 
{\em Decoherence as Decay of the Loschmidt Echo in a Lorentz Gas} Phys Rev. E {\bf 65} 045206(R)\\
Wisniacki D A, Vergini E G, Pastawski H M and Cucchietti F M 2002 
{\em Sensitivity to perturbations in a quantum chaotic billiard} Phys. Rev. E {\bf 65} 055206(R)\\
Cucchietti F M, Lewenkopf C H, Mucciolo E R, Pastawski H M and Vallejos R O 2001
{\em Measuring the Lyapunov exponent using quantum mechanics}, preprint {\tt nlin.CD/0112015}\\
Benenti G and Casati G 2002 {\em Sensitivity of Quantum Motion for Classically Chaotic Systems} Phys. Rev. E {\bf 65} 066205\\
Benenti G, Casati G and Veble G 2002 
{\em Asymptotic decay of the classical Loschmidt echo in chaotic systems}, preprint {\tt nlin.CD/0208003}\\
Jacquod Ph, Adagideli I and Beenakker C W J 2002 
{\em Decay of the Loschmidt Echo for quantum states with sub-Planck scale structure}, 
Phys. Rev. Lett. {\bf 89} 154103\\
Wisniacki D A and Cohen D 2001 {\em Quantum irreversibility, perturbation independent decay, and the parametric theory of the local
density of states}, preprint {\tt quant-ph/0111125}\\
Emerson J, Weinstein Y S, Lloyd S and Cory D 2002 
{\em Fidelity Decay as an Efficient Indicator of Quantum Chaos}, preprint {\tt quant-ph/0207099}\\
Wang W and Li B 2002 
{\em Crossover of quantum Loschmidt echo from golden rule decay to perturbation-independent decay},
preprint {\tt nlin.CD/0208013}

\bibitem{Prosen01} Prosen T 2002 {\it On general relation between quantum ergodicity and fidelity of quantum dynamics} Phys.~Rev.~E {\bf 65} 036208  

\bibitem{Ktop} Prosen T and \v Znidari\v c M 2002 {\it Stability of quantum motion and correlation decay} J.~Phys.~A {\bf 35} 1455

\bibitem{Decoh} Prosen T, Seligman T~H and \v Znidari\v c M 2002 {\it Stability of quantum coherence and correlation decay}, preprint {\tt quant-ph/0204043}

\bibitem{Purfid} Prosen T and Seligman T~H 2002 {\it Decoherence of spin echoes} J.~Phys.~A {\bf 35} 4707

\bibitem{QC} Prosen T and \v Znidari\v c M 2001 {\it Can quantum chaos enhance stability of quantum computation?} J.~Phys.~A {\bf 34} L681

\bibitem{Miller} Miller P~A and Sarkar S 1999 {\it Signatures of chaos in the entanglement of two coupled kicked tops} Phys.~Rev.~E {\bf 60} 1542--50

\bibitem{Uhlmann76} Uhlmann A 1976 {\it The transition probability in the state space of a $A^*$-algebra} Rep.~Math.~Phys. {\bf 9} 273

\bibitem{Haake1} Haake F 1991 {\em Quantum signatures of chaos} (Springer, Berlin)\\ 
Haake F, Ku\'s M and Scharf R 1987 {\it } Z.~Phys.~B {\bf 65} 381

\bibitem{Haake&Strunz} Braun D, Haake F and Strunz W~T 2001 {\it Universality of decoherence} Phys.~Rev.~Lett. {\bf 86} 2913\\ 
Strunz W~Z, Haake F and Braun D 2002 {\it Universality of decoherence in the macroworld}, preprint {\tt quant-ph/0204129}\\ 
Strunz W~T and Haake F 2002 {\it Decoherence scenarios from micro- to macroscopic superpositions}, preprint {\tt quant-ph/0205108} 

\bibitem{Kolovsky} Kolovsky A~R 1994 {\it Number of degrees of freedom for a thermostat} Phys.~Rev.~E {\bf 50} 3569

\bibitem{Qoptics} Meystre P and Sargent M 1990 {\it Elements of Quantum Optics} (Springer, Berlin-Heidelberg)

\bibitem{haroche} J.~M.~Raimond, M.~Brune and S.~Haroche 2001 {\em Colloquium: Manipulating quantum entanglement with atoms and photons in a cavity} Rev.~Mod.~Phys. {\bf 73} 565

\bibitem{Gorin} Gorin T and Seligman T H 2001 {\it Decoherence in chaotic and integrable systems: A random matrix approach}, preprint {\tt nlin.CD/0101018}\\ 
Gorin T and Seligman T H 2001 {\it A random matrix approach to decoherence}, preprint {\tt quant-ph/0112030}

\bibitem{Tanaka} Tanaka A, Fujisaki H and Miyadera T 2002 {\it Saturation of the
prediction of quantum entanglement between weakly coupled mapping systems in strongly
chaotic region}, preprint {\tt quant-ph/0209086}

\end{thebibliography}
\end{document}